\documentclass[twocolumn,prl]{revtex4}

\usepackage[T1]{fontenc}
\usepackage{bm}
\usepackage{graphicx}
\usepackage{epstopdf}
\usepackage{amsmath}
\usepackage{color}
\usepackage[normalem]{ulem}
\usepackage{braket}
\usepackage{graphicx}
\usepackage{amsmath}
\usepackage[T1]{fontenc}
\usepackage{color}
\def\beq{\begin{equation}}
\def\eeq{\end{equation}}
\def\ee{{\rm e}}

\begin{document}

\title{Doubly dressed bosons -- exciton-polaritons in a strong terahertz field}

\author{B.~Pi\k{e}tka$^{1}$} \thanks{Barbara.Pietka@fuw.edu.pl}
\author{N.~Bobrovska$^{2}$}
\author{D.~Stephan$^{1,3}$}
\author{M.~Teich$^{3}$}
\author{M.~Kr\'{o}l$^{1}$}
\author{S.~Winnerl$^{3}$}
\author{A.~Pashkin$^{3}$}
\author{R.~Mirek$^{1}$}
\author{K.~Lekenta$^{1}$}
\author{F.~Morier-Genoud$^{4}$}
\author{H.~Schneider$^{3}$}
\author{B.~Deveaud$^{4,5}$}
\author{M.~Helm$^{3}$}
\author{M.~Matuszewski$^{2}$}
\author{J.~Szczytko$^{1}$}

\affiliation{$^{1}$Institute of Experimental Physics, Faculty of Physics, University of Warsaw, ul. Pasteura 5, 02-093 Warsaw, Poland}
\affiliation{$^{2}$The Institute of Physics, Polish Academy of Sciences, al. Lotnikow 32/46, 02-668 Warsaw, Poland}
\affiliation{ $^{3}$Institute of Ion Beam Physics and Materials Research, HZDR, P.O. Box 510119, 01314 Dresden Dresden, Germany}
\affiliation{$^{4}$Institute of Physics, Ecole Polytechnique F\'ed\'erale de Lausanne (EPFL), Station 3, 1015 Lausanne, Switzerland}
\affiliation{$^5$Ecole Polytechnique, F-91128 Palaiseau, France}

\begin{abstract}
We demonstrate the existence of a novel quasiparticle: an exciton in a semiconductor doubly dressed with two photons of different wavelengths: near infrared cavity photon and terahertz (THz) photon, with the THz coupling strength approaching the ultra-strong coupling regime. This quasiparticle is composed of three different bosons, being a mixture of a matter-light quasiparticle. Our observations are confirmed by a detailed theoretical analysis, treating quantum mechanically all three bosonic fields. The doubly dressed quasiparticles retain the bosonic nature of their constituents, but their internal quantum structure strongly depends on the intensity of the applied terahertz field. 
\end{abstract}


\maketitle

The research on light -- matter interaction in the strong coupling regime, when quantum light emitter and photons can coherently exchange energy, before the coherence is lost, is one of the fundamental problems in cavity quantum electrodynamics (QED). This problem was widely adopted in an atom -- cavity system, described within renowned Jaynes-Cummings model and even beyond this limit, in the ultra-strong coupling regime. The system based on exciton-polaritons, quasiparticles composed from excitons in a semiconductor strongly coupled to the vacuum light field in the cavity, has many advantages. These quantum light emitters do not show fermion-like statistics, but are truly designed bosons that can exhibit non-equilibrium Bose-Einstein phase transition \cite{Kasprzak}. In the present manuscript, we study the phenomenon of double dressing: an exciton coupled to two photonic fields from distinct energy ranges: near-infrared (NIR) and terahertz (THz), which bares no direct analogy in atomic physics. 

Excitons, or bound electron-hole pairs, can be created by the absorption of a near infrared (NIR) photon in a semiconductor. The strong coupling~\cite{ScullyZubairy, Haroche,QED} of excitons and photons in a high-quality microcavity structure results in the formation of exciton-polaritons, due to vacuum field Rabi coupling, evidenced by the appearance of lower polariton (LP) and upper polariton (UP)  resonances~\cite{weisbuch}. The strong coupling regime can be achieved when the coupling strength is larger than the decoherence and the correct description of the system is in terms of new quantum eigenstates, or dressed quasiparticles. In a quantum well (QW), which confines the exciton into a plane, the internal structure of the exciton resembles that of a two-dimensional hydrogen atom. The transitions between the internal states lie in the THz range of the electromagnetic spectrum.
The possibility to induce such transitions with THz photons was shown in Refs.~\onlinecite{n1, n2, n3}. Upon intense THz illumination, close to the 1$s$-2$p$ excitonic transition, Autler-Townes splitting of excitonic states has been observed~\cite{wagner, teich}.

In this letter we demonstrate the simultaneous dressing of excitons with NIR and THz photons. We observe the appearance of a third dressed polariton mode, the middle polariton (MP), which is accompanied by an energy shift of the upper and lower polariton states. We describe our observations with a quantum model that takes into account the relevant couplings between four bosonic fields, including 1$s$ and 2$p$ excitons together with NIR and THz photons. Previously, doubly dressed states of a two-level quantum dot system with two optical photons~\cite{He_DoublyDressed} were shown, as well as interaction of exciton-polaritons with THz photons~\cite{Tomaino}. However, quantum dots exhibit fermionic-like single-photon emission and photon anti-bunching. By contrast, exciton-polaritons are bosonic particles, which also applies to our new quasi-particle observed here.

\begin{figure}
	\centering
	\includegraphics[width=.36\textwidth]{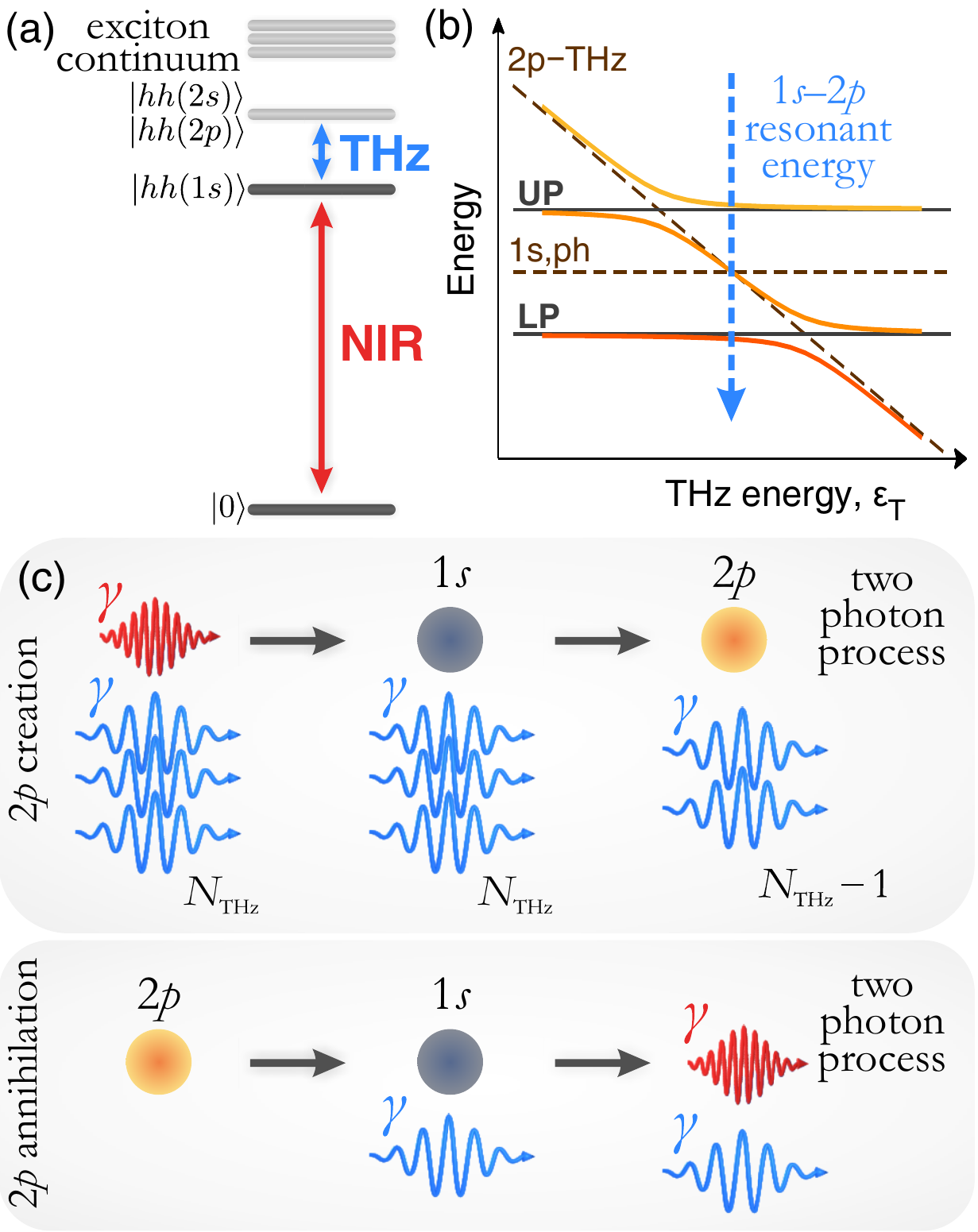}
	\caption{a)~Scheme of the internal structure of an atom-like exciton. The vacuum state ($\ket{0}$) is coupled to a ground 1$s$ heavy hole (hh) exciton via the NIR photon. The 1$s$ is coupled to the excited 2$p$ exciton via THz photon. In our theoretical model we neglect other excited electronic states, but we include the exciton ionization and blueshift from the dynamical Franz-Keldysh effect. b)~The bare exciton-polariton energies (black solid lines) and the double dressed quasiparticle energy diagram (yellow-orange lines). The THz energy resonant to the 1$s$--2$p$ transition is marked with a blue arrow. c)~Scheme of the creation and annihilation of the 2$p$ exciton state. This process involves two photons, one from NIR and one from THz spectral ranges.}
	\label{f1}
\end{figure}


To confine the NIR photonic mode we used a GaAs lambda microcavity sandwiched between two AlAs/GaAs distributed Bragg reflectors. A single 8~nm-thick In$_{0.04}$Ga$_{0.96}$As quantum well was placed, at the maximum of the cavity field, providing the excitonic component of polaritons. Excitons confined inside the quantum well couple to the photon modes with a coupling strength given by the Rabi splitting, $\Omega_{C}$. In the case of our sample, the excitonic resonance was at approximately~1.484~eV and the vacuum Rabi splitting was equal to approximately~$\Omega_{C}$~=~3.5~meV. The on-resonance polariton linewidth was 0.3~meV allowing for a clear resolution of polariton states. Details on the sample structure and optical characterization can be found in Ref.~\onlinecite{ounsi} and Ref.~\onlinecite{BenoitPietka}, respectively. The sample was kept in a cryostat at the temperature of 6 K.
The polariton population was created resonantly by fs laser pulses, with the spot size of 150~$\mu$m. The pulse spectrum was sufficiently broad to simultaneously excite LP and UP branches at zero momentum. In the experiments reported here, the excitation power was low enough to maintain polaritons in the linear regime of polariton-polariton interactions. All experimental observations are reported for close to zero exciton-cavity photon detuning $\delta~\in$~(-0.2,0.2)~meV.

The THz light was generated at the Helmholtz-Zentrum Dresden-Rossendorf by the free-electron laser (FEL). The spectral width of the pulse at the resonant wavelength of 182~$\mu$m was 1.6~$\mu$m (which corresponds to (6.80~$\pm$~0.03)~meV). The THz pulse duration was determined to be approx. 30~ps~\cite{teich}, and the spot size was about 1~mm in diameter. Since this is significantly larger than the NIR spot size, the THz intensity can be regarded to be uniform inside the probed area. The scheme of the experimental setup is provided in the Supplementary Material (SI).

The exact energy separation between the 1$s$ and 2$s$ exciton states was determined experimentally~\cite{2s}. We have observed the appearance of the 2$s$ exciton-polariton in photoluminescence spectra in magnetic fields.
From the observed energy change in magnetic fields we could determine the zero field energy splitting between 1$s$ and 2$s$ exciton in our sample to approx. 6.5 meV. Moreover, at zero magnetic field 2$s$ and 2$p$ states are very close in energy and we can estimate the energy difference between 1$s$ and 2$p$ state to $6.5\pm0.5$~meV. Our measurement results agree with other works performed on similar structures \cite{teich}.


\begin{figure}
	\centering
	\includegraphics[width=.47\textwidth]{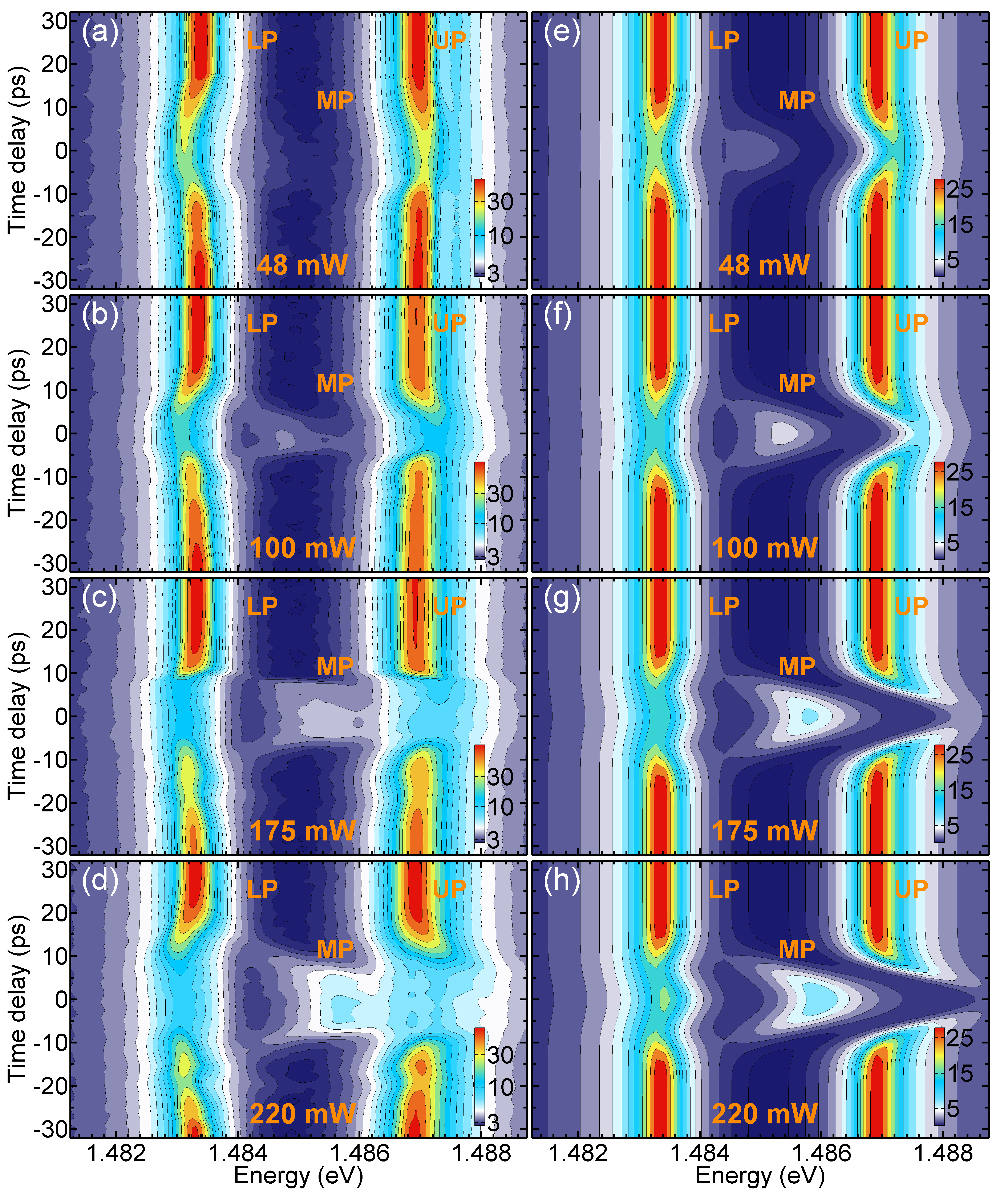} 
	\caption{Transmission spectra of lower (LP) and upper (UP) polaritons (at normal incidence) as a function of the delay between the NIR and THz pulses (in equidistant time intervals). The color scale indicates the intensity in arbitrary units. The THz photon energy is 6.8~meV, slightly detuned from 1$s$--2$p$ excitonic transition ($6.5\pm0.5$~meV). The appearance of THz induced middle branch (MP) at the maximum of THz pulse (zero delay) is visible. This is accompanied by an energy shift of LP and UP branches. a)- d) Illustrates the behavior with increasing average THz power as marked directly in the image. Positive delays correspond to the NIR pulse preceding the THz pulse, negative delays to the reversed situation. e) - h) Corresponding theoretical model. The color scale is nonlinear to enhance the low-intensity signal at zero delay time.}
	\label{f2}
\end{figure}

\begin{figure}
	\includegraphics[width=.45\textwidth]{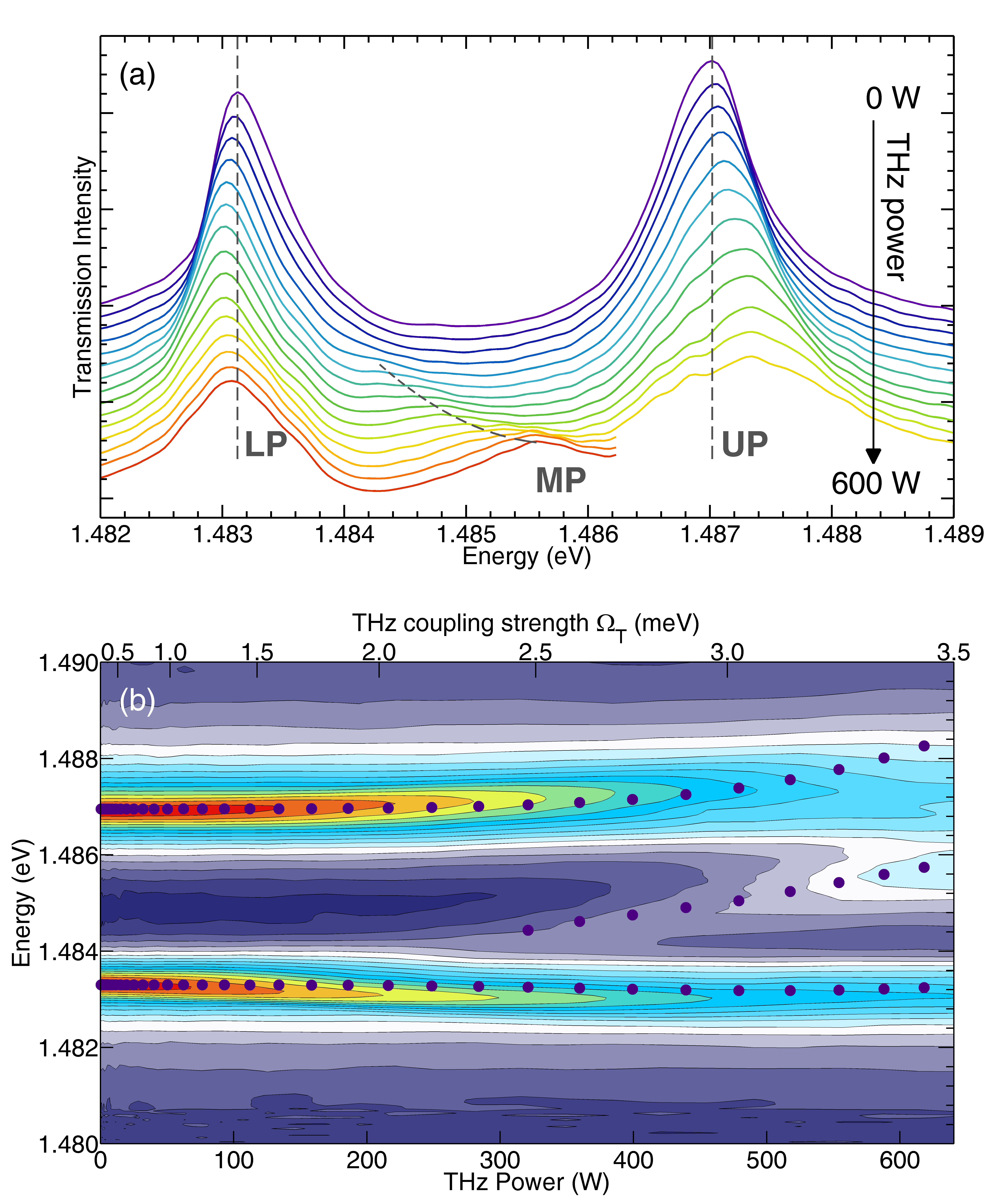} 
	\caption{Transmission spectra (at normal incidence) of semiconductor microcavity revealing dressed polaritons and approaching to ultra-strong coupling regime. a) Spectra illustrate the cross section of the results shown in Fig.~\ref{f2}d for the time delays from 15~ps to 0~ps. At weak perturbation (at 15~ps time delay, that corresponds to the onset of THz pulse) the transmission spectra reveals unperturbed LP and UP states. As THz intensity increases, a red-shift of the LP and a blue-shift of the UP are observed and the additional MP line is visible (the dotted lines are guides to eye to compare the energy shifts, for vertical lines the energy is constant, the middle dotted line follows the MP energy shift). The THz induced opacity of the sample is revealed by the loss of the overall intensity of the transmitted signal. b) Comparison of the experimental data from Fig.~\ref{f2}d (colored map) with theoretical model from Fig.~\ref{f2}h (dots). The time delay can be converted to instantaneous THz intensity, as indicated directly in the image (details can be found in the SI). The interaction energy with the THz field is illustrated in the top scale.}
	\label{f3}
\end{figure}

The idea of the experiment is as follows: a weak NIR laser pulse creates a population of exciton polaritons through the excitation of an initially empty semiconductor QW inside a microcavity (Fig.~\ref{f1}a). A strong THz external laser pulse, tuned to resonance with the 1$s$ to 2$p$ transition, mixes the two excitonic states, as shown in Fig.~\ref{f1}b. Fig.~\ref{f1}c illustrates the 2$p$ excitation and decay scheme. The two external laser beams are collinear and synchronized. The NIR Ti:Sapphire laser provides 12~fs pulses, with central wavelength tuned to the excitonic resonance. The spectral width was broad enough to cover both LP and UP. Therefore exciton-polaritons were created resonantly in a superposition of polariton states.

The NIR transmission spectra taken for normal incidence as a function of the delay time between the NIR and THz pulses are reported in Fig.~\ref{f2}. When the system is unperturbed by the THz beam, i.e. at large positive or negative delays, we observe the two expected transmission resonances corresponding to LP and UP polaritons as in Ref.~\onlinecite{BenoitPietka} and in Supplemental Information (SI). For delays of the order of 15~ps, the THz pulse intensity starts to increase and the energy and the intensity of both polariton lines are modified. The energy of LP and UP are repelled apart, which is visible already at low average THz power, Fig.~\ref{f2}(a,e). At slightly higher average THz intensity, Fig.~\ref{f2}(b,f), the LP and UP energy shift is faster and one additional line appears between polariton states, getting stronger for larger THz intensities, Fig.~\ref{f2}(c,g) and (d,h). We will refer to this line as the middle polariton branch (MP).

In all cases, the LP and UP energy shifts and the appearance of the MP are accompanied by a drastic decrease in the signal intensity, especially pronounced for exact time overlap between the pulses where the sample is getting almost opaque. A THz average power as high as  200~mW gives an electric field strength of 12~kV$\cdot$~cm$^{-1}$ (or intensity of 1~MW$\cdot$~cm$^{-2}$) at the position of the QW. In such a regime of THz excitation the excitons can ionize. 

The appearance of the middle polariton line should in no case be considered as an indication of the transition to the weak coupling regime~\cite{Takemura} and the saturation of the exciton transition. This is confirmed by the following: a) all three lines coexist at a given time and THz intensity; b) simultaneous energy shift of LP, UP and MP with increasing THz intensity (the energy of cavity photon is independent on the THz intensity); c) the MP energy depends on the THz power (Fig.~\ref{f3}a) and THz detuning from 1$s$-2$p$ transition (see SI). All images illustrated in Fig.~\ref{f2} are symmetric versus zero delay time, which demonstrates that even at very high THz powers the exciton population is not destroyed. Indeed, we are in the low NIR excitation regime and the total excited exciton population is far from the saturation density.

To understand the physics behind our observations, we performed calculations based on the four-mode Hamiltonian. The derivation (described in the SI) is based on the rotating wave approximation, focusing on the four resonantly interacting bosonic modes. A 1$s$ exciton can be formed by absorption of a NIR photon. The 1$s$ exciton can further absorb a photon from the external THz field to form an excited 2$p$ state, see Fig.~\ref{f1}c. The annihilation process of a 2$p$ exciton is a reverse two photon process. These are described by 

\begin{align}
H &= \sum_{\nu=1s,2p,C,T} \epsilon_\nu \hat{a}_\nu^\dagger \hat{a}_\nu + \hbar \Big(g_C \hat{a}_C \hat{a}_{1s}^\dagger + g_T \hat{a}_T \hat{a}_{1s} \hat{a}_{2p}^\dagger + \nonumber\\
&+g_C \hat{a}_C^\dagger \hat{a}_{1s} + g_T \hat{a}_T^\dagger \hat{a}_{1s}^\dagger \hat{a}_{2p} \Big) + H_{\rm high}.
\label{H_full_RWA}
\end{align}

Here, $\hat{a}_{1s},\hat{a}_{2p},\hat{a}_{C}$, and $\hat{a}_{T}$ are bosonic annihilation operators for the 1$s$ and 2$p$ excitons, cavity and THz photons, and $\epsilon_{1s,2p,C,T}$ are their respective energies. The coupling of the 1$s$ excitons to photons is given by $g_C$, while $g_T$ determines the probability of the 1$s$--2$p$  transition with simultaneous annihilation of a THz photon. The $H_{\rm high}$ term describes coupling to other electronic excitations.

This Hamiltonian does not provide a simple bosonic quasiparticle spectrum, even when neglecting this last term. However, as shown in the SI, when the occupation of the THz mode is large,
the elementary bosonic excitations (dressed states) turn out to be the solutions of
\begin{equation}
H_{single}=\left(
\begin{array}{ccc}
\epsilon_C & \Omega_C^*/2 & 0 \\
\Omega_C/2 & \epsilon_{1s} & \Omega_T^*/2 \\
0 & \Omega_T/2 & \epsilon_{2p}-\epsilon_T \label{H_single}
\end{array}
\right)
\end{equation}
where $\Omega_C=2\hbar g_C$, $\Omega_T=2\hbar g_T \sqrt{N_T}$, with $N_T$ the average occupation of the THz laser mode,
$\Delta=\epsilon_{2p}-\epsilon_{1s}-\epsilon_T$ is the detuning of the THz field,
and $\delta=\epsilon_C-\epsilon_{1s}$ is the exciton-photon detuning.
Note that the 1$s$ exciton-cavity coupling $\Omega_C$ is independent
of the number of photons, and corresponds to the vacuum Rabi splitting~\cite{Snoke_VacuumRabi}. In contrast,
the 1$s$--2$p$ coupling $\Omega_T$ is proportional to the square root of the number of photons in the coupling THz field, which is in analogy with the
Autler-Townes effect~\cite{CohenTannoudji_AutlerTownes}.
The above matrix has three eigenstates, which correspond to lower, middle, and upper polaritons, as illustrated in Fig.~\ref{f1}d.
In the absence of a strong THz field, one of the modes corresponds to excitation of the uncoupled, optically
inactive bare 2$p$ exciton.

We reproduce the experimental spectra by computing the solutions of~(\ref{H_single}), while including pumping and losses (see SI). The effects coming from the higher electronic states $H_{\rm high}$ can be included, with a good accuracy, when considering two phenomena. The effect of exciton ionization~\cite{Miller_ExcitonIonization,Koch_Ionization} by the strong electric field of the THz laser leads to the decay of the LP and UP lines at high values of the THz intensity, as visible in Fig.~\ref{f2}. At the same time, the dynamical Franz-Keldysh effect increases the energy of the exciton states due to the induced motion of electron and hole, which is most clearly seen in the asymmetry of the LP and UP energy shifts (Fig.~\ref{f3}).

Fig.~\ref{f3}a illustrates a few NIR transmission spectra that correspond to the cross sections of the map from Fig.~\ref{f2}d. Fig.~\ref{f3}b  shows the exact comparison of the experimental data (Fig.~\ref{f2}d) with the theoretical model (Fig.~\ref{f2}h). The time-averaged THz pulse power as high as 200~mW corresponds to $N_{T} = 1.4\times 10^{13}$ photons in a single coherent pulse. As the THz coupling strength, $\Omega_T \propto \sqrt{N_T}$, the dipole interaction energy (top scale in Fig.~\ref{f3}b) can approach the same order as the transition energy (approx. 6.5~meV). Therefore, we can tune the strength of the interactions through all regimes, from weak to ultra strong~\cite{giacomo}.

In conclusion, strong coupling involving both the vacuum field Rabi splitting and the coupling to a second propagating field gives rise to a novel class of coherent phenomena in the solid-state, where the open dissipative character of the system competes with strong coherence. Bosons doubly dressed with optical and THz fields are of fundamental interest due to the nontrivial internal structure of the quantum state. Our states are superpositions of NIR photons, THz photons and excitons. We have shown here that the dressed particle description remains valid, albeit with the addition of a third dressed state in the picture. The coupling to a strong coherent THz field might be considered in the design of semiconductor-based THz devices \cite{KavokinTHz1, KavokinTHz2}. Strong opacity of the micro-cavity structure induced by the THz beam allows one to consider possible applications such as ultrafast optical switches.

We acknowledge the support from the National Science Center (NCN) grant 2011/01/D/ST7/04088. NB and MM acknowledge the support from the NCN grant 2011/01/D/ST3/00482. Part of this work has been supported by CALIPSO under the EC contract 312284.

\bibliography{bibliography.bib}

\clearpage

\thispagestyle{empty}

\section{Supplementary information for ``Doubly dressed bosons -- exciton-polaritons in a strong terahertz field''}


\section{Experimental details}

\subsection{Exciton-polariton concept}

The exciton-photon coupled state in the NIR range is assured by the sample structure. This technically challenging task is realized by placing an optically active quantum well at the antinode of the electric field distribution between two Bragg mirrors of the microcavity.

\begin{figure}[h]
\includegraphics[width=6cm]{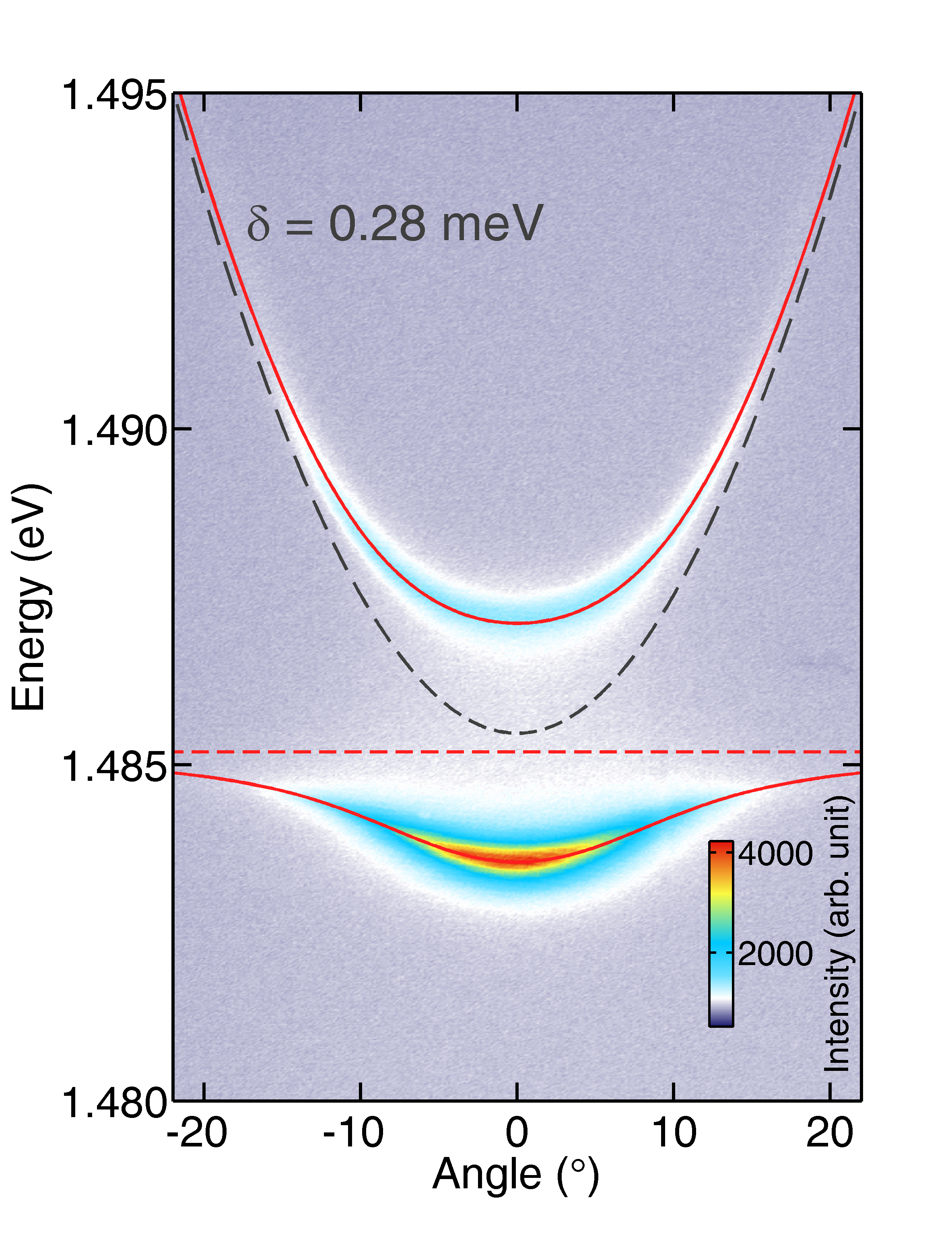}
\caption{Typical angularly resolved photoluminescence spectrum of the sample taken at the position of the exciton-photon detuning of 0.28meV. The detuning is determined as the energy difference between bare exciton ($\epsilon_{1s}$) and photon ($\epsilon_C$) energies,  $\delta=\epsilon_C-\epsilon_{1s}$. Two observed resonances are LP and UP modes. The photoluminescence signal was excited non-resonantly with a ps laser pulses. The energy of the excitation corresponds to the first reflectivity minima of the cavity from high energy side. The energy of LP and UP modes is marked with red solid lines and bare exciton and photon modes with red and black dashed line, respectively.}
\label{fig:PL_spectrum}
\end{figure}

The sample under investigation consists of an 8 nm thick single GaInAs quantum well sandwiched between two AlAs/GaAs distributed Bragg reflectors (DBR) forming a lambda microcavity. The microcavity is wedged and the exciton-photon detuning can be changed by changing the position of the excitation spot on the sample. The excitonic resonance is at 835 nm and the vacuum Rabi splitting is $\Omega_{C}$  = 3.5 meV. The typical angle-resolved photoluminescence spectrum of the sample is illustrated in Fig.~\ref{fig:PL_spectrum}. The strong coupling regime is observed: excitons and photons are mixed, as they form two new eigenstates of the system, called lower (LP) and upper (UP) polaritons.

 These bosonic quasiparticles have a dual nature, inheriting properties of both of its constituents: a small mass from the photon part (5 order of magnitude smaller than free electron mass) and the possibility to interact, through Coulomb scattering, from the excitonic part. At zero detuning, the energy separation between the LP and UP states is given by the vacuum  field Rabi splitting, $\Omega_{C}$. 
 
The exciton-polariton dressed states can be further modified by an external, resonant and strong THz field. Exciton-polaritons inherit their excited-shell structure from the excitonic component. Therefore exciton-polaritons are sensitive to low-energy excitations.  If a sufficiently narrowband, external THz laser is tuned close to one of the allowed transitions within the excitonic excited shell structure, the exciton-polariton can be dressed with this photon field. The first excitation, 1$s$ to 2$p$ transition, lies in the THz regime. In our sample, the resonant energy was estimated, through magneto-spectroscopy to be $6.5\pm0.5$~meV.

\subsection{Setup}

\begin{figure}[h]
\includegraphics[width=8cm]{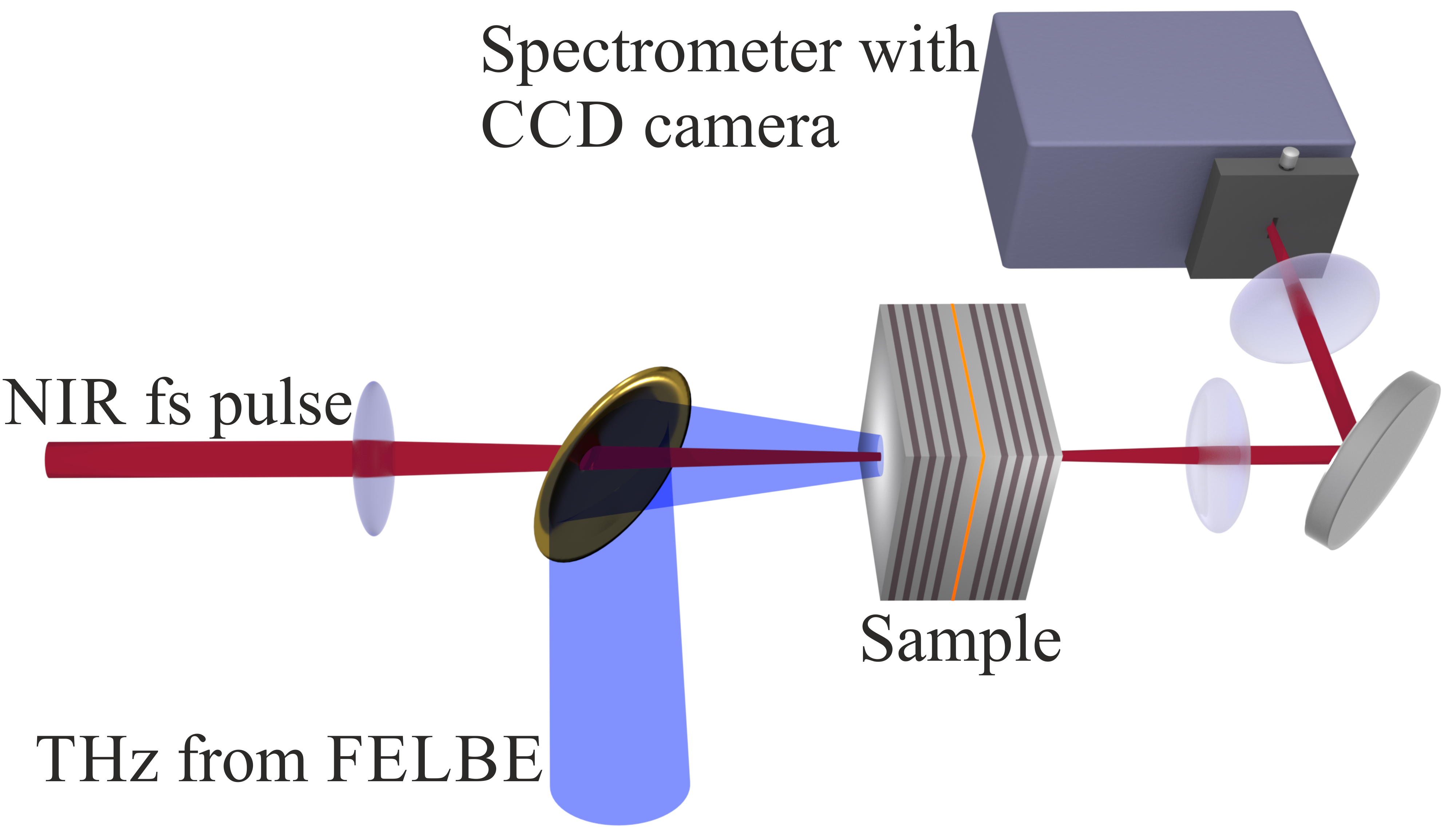}
\caption{Scheme of the experimental setup.}
\label{fig:setup_scheme}
\end{figure}

The scheme of the experimental setup is illustrated in  Fig.~\ref{fig:setup_scheme}. The NIR pulse allowing to measure the transmission of the structure was focused onto the sample with a single lens of long focal length. This allows one to excite polaritons at near zero angle, which corresponds to the creation of polaritons with zero in-plane momentum, at the minimum of their dispersion branch. The transmitted NIR signal was collected with a single lens and propagated in free space to the entrance slit of a spectrometer.
The THz laser beam was focused on the front-side of the sample with a parabolic mirror.  The parabolic mirror had a hole at its centre in order to allow the NIR laser to reach the sample and therefore permitting the simultaneous excitation of the sample by both lasers.  The repetition rate of the NIR-fs laser was reduced by an acousto-optic pulse picker in order to match the 13 MHz  repetition rate of the FEL. Both pulses were synchronized in time with a timing jitter of 1-2~ps and their respective delay time was adjusted using a phase shifter. The sample was kept in a cryostat at the temperature of about 6 K.

In all spectra illustrated in the main paper (Fig.~2 a-d) there is a background signal due to the imperfect blocking of the NIR laser pulses by the pulse-picker that synchronizes NIR and THz lasers. These weak NIR pulses produce a background of unshifted LP and UP spectral lines. This signal is especially visible at zero delay time at high THz powers, Fig.~2 c-d, where its intensity becomes comparable to the signal of doubly dressed quasi-particles. The theoretical images presented in Fig.~2 e-h are free of this background signal.

\begin{figure}
\includegraphics[width=8cm]{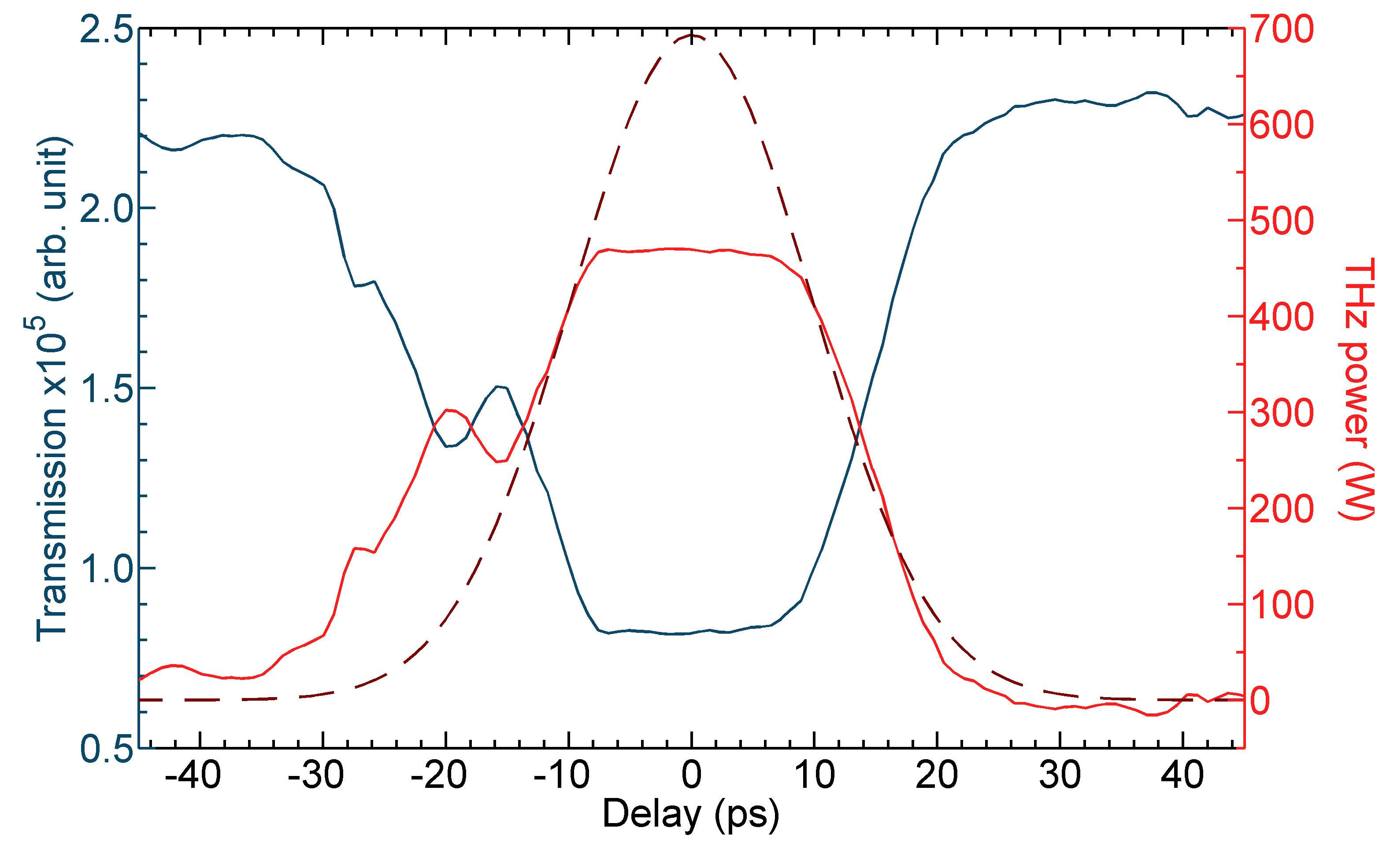}
\caption{NIR transmission signal (energy integrated) for different delay between NIR and THz pulses for average THz intensity of 220 mW (blue curve) and inverse of this signal (red curve), that reflects the instantaneous THz intensity in a pulse. Gaussian fit of the expected intensity in a THz pulse is plotted with red dashed curve.}
\label{fig:beam}
\end{figure}

The THz laser pulses were characterized in detail in Ref.~\onlinecite{teich}. The pulse duration was determined to be approx. 30~ps. The intensity versus time profile of the THz pulse is also directly visible in our experimental data. The NIR signal intensity decreases as the THz intensity increases. The energy integrated NIR transmission signal is plotted in Fig.~\ref{fig:beam} (blue curve). The flat minimum illustrates almost complete quenching of the signal due to the exciton ionization effects. The inverse of this signal (red curve) directly illustrates the THz pulse profile. Assuming the profile to be almost Gaussian (red dashed curve) and a 13 MHz repetition rate of the free electron laser, the time delay between the NIR and THz pulses can be recalculated into the instantaneous THz intensity in the pulse, what is marked by the right scale in Fig.~\ref{fig:beam}.

\begin{figure*}
\includegraphics[width=17cm]{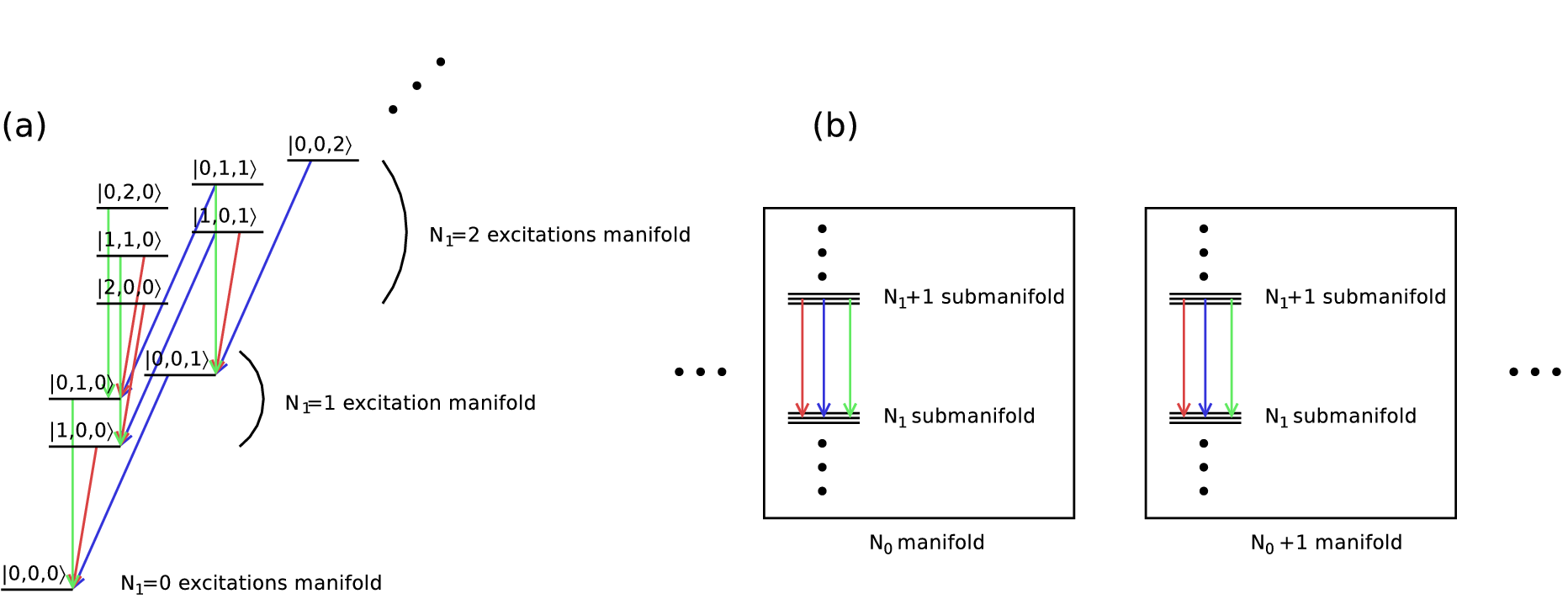}
\caption{Diagrams of energy levels and optical transitions within (a) semiclassical treatment and (b) full quantum treatment.}
\label{fig:diagrams}
\end{figure*}

\section{Quantum model for exciton-polaritons in a strong terahertz field}

\subsection{Hamiltonian}

We consider the interaction of a single optical mode and a single terahertz mode with quantum well excitons, neglecting other
electronic excitations. We first consider the conservative limit, in which we can obtain analytical results without invoking
the classical approximation. The effect of losses will be postponed to the last section.
The fully quantum Hamiltonian in the Coulomb gauge is~\cite{ScullyZubairy, Mahan}
\beq
\hat{H} = \frac{1}{2m_e}\sum_i \left({\bf \hat{p}}_i - e {\bf \hat{A}}\right)^2 + \hat{H}_{\rm el} + \sum_{\nu=T,C} \epsilon_\nu \hat{a}_\nu^\dagger \hat{a}_\nu,
\label{H_full}
\eeq
where ${\bf \hat{p}_i}$ are the momentum operators for electrons,
$\hat{H}_{\rm el}$ is the part of the Hamiltonian corresponding to interactions within the crystal,
$\hat{a}_T,\hat{a}_C$ are annihilation operators for the terahertz mode and the cavity mode,
${\bf \hat{A}}=\vec{\xi}_C  \mathcal{A}_C  \ee^{-i \omega_C t + i{\bf k}_C {\bf r}} \hat{a}_{C} + \vec{\xi}_T \mathcal{A}_T \ee^{-i \omega_T t + i{\bf k}_T {\bf r}} \hat{a}_{T} + {\rm h. c.}$
is the operator of the vector potential,
with $\mathcal{A}_i=(\hbar/2\varepsilon_0 V_i \omega_i)^{1/2}$  and $V_i$ being the appropriate quantization volume.
We limit ourselves to two photonic modes $\hat{a}_T,\hat{a}_C$ with ${\bf k}_i$ vectors perpendicular to the quantum well.
The unit vectors $\vec{\xi}_{C,T}$ describe their polarization.
The photon fields span over the cavity ($\hat{a}_C$) and beyond it ($\hat{a}_T$), but we assume that considerable
matter-light interaction occurs only in the quantum well.
Here we neglect the influence of the external weak NIR probe on the mode structure. The probe would correspond to another
optical mode with small coupling to the cavity mode $\hat{a}_C$~\cite{ScullyZubairy} and no coupling to excitons.

We now make use of the rotating wave approximation (RWA), neglecting the energy-nonconserving,
or non-resonant parts in the Hamiltonian (\ref{H_full}).
The standard form of RWA disregards terms that correspond to coupling with negative
frequencies. Additionally, we assume that only the transitions almost exactly
resonant with the photonic modes frequencies,
that is the 1$s$ exciton creation transition and the 1$s$ to 2$p$ exciton transition, are important. The other transitions will only be
taken into account in the value of the dielectric constant, and later by inclusion of the dynamical Franz-Keldysh effect and exciton ionization. We may call this approximation the
three-level approximation, although it still takes into account an infinite number of energy levels (see Fig.~\ref{fig:diagrams}),
since we maintain the bosonic character of all the fields.
In the low exciton density limit and within the dipole approximation, the Hamiltonian becomes
\begin{align}
  \hat{H} &= \sum_{\nu=T,C,1s,2p} \epsilon_\nu \hat{a}_\nu^\dagger \hat{a}_\nu + \label{H_full_RWA}\\ \nonumber
  &+\hbar \left(g_C \hat{a}_C \hat{a}_{1s}^\dagger + g_T \hat{a}_T \hat{a}_{1s} \hat{a}_{2p}^\dagger +  {\rm h.c.}\right),
\end{align}
where $\hat{a}_{1s},\hat{a}_{2p}$ are annihilation operators for the 1$s$ and 2$p$ excitons.
These operators are linear combinations of products of electron and hole operators~\cite{Elliot,Savona,Kavokin_Microcavities}.
In the low exciton density limit considered here they obey bosonic commutation relations.
Any interactions between the excitons are neglected.
The light-matter couplings in~(\ref{H_full_RWA}) are $g_C=({\bf d}_{1s}\cdot \vec{\xi}_{C})\mathcal{A}_C\omega_C/\varepsilon\hbar$ and
$g_T=({\bf d}_{1s\rightarrow 2p}\cdot \vec{\xi}_{T})\mathcal{A}_T\omega_T/\varepsilon\hbar$,
where ${\bf d}_{1s\rightarrow 2p}=e\langle 2p|{\bf r}|1s\rangle$ is the 1$s$ to 2$p$ transition dipole moment,
and ${\bf d}_{1s}\sim{\bf d}_{vc} |\phi_{1s}(0)|$.
Here $|\phi_{1s}(0)|^2$ is proportional to the probability of finding an electron and hole in the same crystal cell,
and ${\bf d}_{vc}$ is the dipole moment between valence and conduction bands given by their Bloch functions~\cite{Elliot}.
The states $|1s\rangle$ and  $|2p\rangle$ are excited states of the crystal with one exciton present.

For convenience we choose a basis in which $g_C$ and $g_T$ are real.
The exciton modes are localized in the quantum well. In the case of  $N_{\rm QW}$ identical narrow quantum wells,
the exciton modes are delocalized in all wells,
and the exciton transition dipole moment scales as ${\bf d}_{1s}\sim N_{\rm QW}^{1/2}$.

\subsection{Semiclassical treatment}

In the semiclassical approximation, we treat the terahertz laser radiation as a classical field (assuming a coherent state) and replace
the terahertz operator by the c-number $\alpha_T\ee^{-i\omega_T t}$.
This automatically means that we neglect any correlations between the terahertz subspace and the rest (the state is a product state),
which is a rather strong assumption. Nevertheless, we will generalise it in the next subsection to the full quantum treatment.
In the classical version of the RWA we shift the energy of the 2$p$ state to $\epsilon_{2p} \rightarrow \epsilon_{2p} - \epsilon_T$ to obtain
\beq
\hat{H} = \sum_{\nu=C,1s,2p} \epsilon_\nu \hat{a}_\nu^\dagger \hat{a}_\nu
+ \hbar\left(g_C \hat{a}_{C} \hat{a}_{1s}^\dagger
+ g_T \alpha_T \hat{a}_{1s}\hat{a}_{2p}^\dagger  + {\rm h.c.}\right) \label{H_RWA}
\eeq

Let us now consider the manifold where only one excitation (exciton or cavity photon) is present. The Hamiltonian (\ref{H_RWA})
written in the basis of bare states $|N_C, N_{1s}, N_{2p}\rangle=|1,0,0\rangle, |0,1,0\rangle, |0,0,1\rangle$ is
\beq
H_{single}=\left(
\begin{array}{ccc}
\epsilon_C & \Omega_C^*/2 & 0 \\
\Omega_C/2 & \epsilon_{1s} & \Omega_T^*/2 \\
0 & \Omega_T/2 & \epsilon_{2p}-\epsilon_T \label{H_single}
\end{array}
\right)
\eeq
where $\Omega_C=2\hbar g_C$, $\Omega_T=2\hbar \alpha_T g_T$,
$\Delta=\epsilon_{2p}-\epsilon_{1s}-\epsilon_T$ is the detuning of the THz field,
$\delta=\epsilon_C-\epsilon_{1s}$ is the exciton-photon
detuning.
The $\Omega_T\sim \sqrt{N_T}$ depends on the THz field intensity while $\Omega_C$ is fixed.

For the above $3\times 3$ hermitian matrix we can find 3 orthonormal eigenvectors $v_i=(U_{i,C},U_{i,1s},U_{i,2p})^T$
with real eigenvalues $E_i$, where $i=1\dots 3$. The matrix $U_{ij}$ is unitary, $UU^\dagger=1$. Let us define the operators
\beq
\hat{b}_i = U_{i,C} \hat{a}_C + U_{i,1s} \hat{a}_{1s} + U_{i,2p} \hat{a}_{2p} \label{bi}
\eeq
The states $\hat{b}^\dagger_i|0,0,0\rangle$ are exactly the dressed states, since they are the eigenstates of the Hamiltonian
with  $\Omega_C, \Omega_T$ couplings included. In particular,
\beq
\hat{H} |d_i\rangle = E_i |d_i\rangle, \qquad {\rm where}\quad  |d_i\rangle = \hat{b}_i^\dagger|0,0,0\rangle, i=1\dots 3. \label{dressed}
\eeq
The inverse relations for the operators $\hat{a}_c$, $\hat{a}_{1s}$, and $\hat{a}_{2p}$ are given by the elements of
the inverse matrix
\beq
\hat{a}_j = \sum_i U_{ji}^\dagger \hat{b}_i, \qquad {\rm where} \quad j=C,1s,2p.
\label{inverse}
\eeq

Now we proceed to include states with multiple excitations. The operators $\hat{b}_i$
possess the right bosonic commutation relations: $[\hat{b}_i,\hat{b}_i^\dagger]=1$,
$[\hat{b}_i,\hat{b}_j^\dagger]=0$ for $i\neq j$, and $[\hat{b}_i,\hat{b}_j]=0$ which follows from the unitarity of $U_{ij}$.
From (\ref{H_RWA}) and (\ref{inverse}) it is clear that the Hamiltonian written in the
basis of the dressed states has the general quadratic form
\beq
\hat{H}= \sum_{ij} H_{ij} \hat{b}_i^\dagger \hat{b}_j
\label{general_form}
\eeq
where $H_{ij}$ are c-numbers. Substituting the above into (\ref{dressed}) for $i=1\dots 3$ and using the
commutation relations it is easy to see that
$H_{ij}=0$ for $i \neq j$. Hence the Hamiltonian is diagonal in the dressed states basis
\beq
\hat{H} = \sum E_i \hat{b}_i^\dagger \hat{b}_i, \quad i=1\dots 3\,.
\eeq
The absorption, emission, reflectance and transmission spectra in the weak-probe limit are related to the matrix elements
$\langle d_i | \hat{a}_C | d_j \rangle$ and $\langle d_i | \hat{a}_C^\dagger | d_j \rangle$
where $| d_{i,j} \rangle$ belong to the neighboring manifolds
(the number of excitations differs by one).
This can be seen easily by adding external optical modes to the Hamiltonian (\ref{H_RWA}), coupled to the
field of the cavity.
Since $\hat{a}_C$ is a linear combination of $\hat{b}_i$, Eq.~(\ref{inverse}), we expect to have
in general 3 spectral lines
with energies equal to the energy of each of the dressed excitations $E_i$.
The relative intensities of the lines are given by their weights
$|U^\dagger_{C,i}|^2=|U_{i,C}|^2$.

\subsection{Full quantum treatment}

Only several modifications to the above derivation are necessary to perform a full quantum treatment.
There are two constants of motion for the above Hamiltonian (\ref{H_full_RWA}),
the number of ``terahertz-induced'' excitations, $N_0=N_T+N_{2p}$ and
the number of cavity photons plus excitons $N_1=N_C+N_{1s}+N_{2p}$.
The latter corresponds to the number of excitations in the semiclassical treatment.
The manifolds of states with the same $N_0$ break up into submanifolds with a given $N_1$.
The operators $\hat{a}_C$ and $\hat{a}_C^\dagger$, related to the optical properties, commute with $N_0=a^\dagger_T a_T + a^\dagger_{2p} a_{2p}$.
Consequently, the optical transitions are between
states from neighboring submanifolds (where the number of excitations $N_1$ differs by one), but from the same $N_0$ manifold.
The $N_0$ manifolds are completely isolated from each other.
Moreover, if the terahertz laser field is close to a coherent state with large number of
quanta, the fluctuations of $N_0$ will be small, $\Delta N_0 / N_0 = O(N_0^{-1/2})$.
We can then treat $N_0$ as a constant. This assumption is less strong
and much more precise than the one made in the semiclassical treatment.

To diagonalize the Hamiltonian, we have to make another assumption: that the number of excitations (polaritons) is much smaller than the number of terahertz quanta: $N_1 \ll N_0$.
This is fulfilled in the case of a strong THz field and a weak probe.
Let us now define the operator
\beq
\hat{B} = \frac{1}{\sqrt{N_0}} a_{2p}a_T^\dagger
\eeq
It is easy to check that $\hat{B}$ possesses the right bosonic commutation relations with the accuracy of the order of $O(N_1/N_0)$. We can now write
\beq
\hat{B}^\dagger \hat{B}= \frac{1}{N_0} a^\dagger_{2p} a_{2p} a_T a_T^\dagger \approx a^\dagger_{2p} a_{2p}
\label{BdB}
\eeq
where the error is again of the order of $O(N_1/N_0)$. This allows to write the Hamiltonian as
\begin{align}
\hat{H} &= \sum_{\nu=C,1s} \epsilon_\nu \hat{a}_\nu^\dagger \hat{a}_\nu + \left(\epsilon_{2p}-\epsilon_{T}\right) \hat{B}^\dagger \hat{B} + \epsilon_T N_0+ \nonumber\\
&+ \hbar\left(g_C \hat{a}_{C} \hat{a}_{1s}^\dagger
+ g_T \sqrt{N_0}\hat{a}_{1s}\hat{B}^\dagger  + {\rm h.c.}\right) \label{H_reduced}
\end{align}
where we have used the relation $\epsilon_{2p} \hat{a}_{2p}^\dagger \hat{a}_{2p} +\epsilon_{T} \hat{a}_{T}^\dagger \hat{a}_{T} =
(\epsilon_{2p} - \epsilon_{T}) \hat{a}_{2p}^\dagger \hat{a}_{2p} + \epsilon_T N_0$. Apart from the constant term,
the form of the above Hamiltonian is the same as (\ref{H_RWA}) provided that the replacements $\hat{a}_{2p}\rightarrow \hat{B}$
and $\alpha_T \rightarrow \sqrt{N_0}$ are made. We can now follow the steps of the semiclassical treatment to
obtain the diagonal form of the Hamiltonian.

\subsection{Classical treatment}
Since we treat excitons as bosonic excitations, it is possible to use a simple classical treatment.
We replace all quantum operators in the
Hamiltonian (\ref{H_full_RWA}) with c-numbers, assuming that all photon and matter fields are
in coherent states. This assumption is justified if the system is excited resonantly by a coherent laser probe and the effects
of decoherence are negligible. We obtain a set of dynamical equations
\beq
\left\{
\begin{array}{l}
i \dot{\alpha}_T = \omega_T \alpha_T + g_T \alpha_{1s}^* \alpha_{2p},\\
i \dot{\alpha}_C = \omega_C \alpha_C + g_C \alpha_{1s},\\
i \dot{\alpha}_{1s} = \omega_{1s} \alpha_{1s} + g_C \alpha_{C} + g_T \alpha_{T}^* \alpha_{2p},\\
i \dot{\alpha}_{2p} = \omega_{2p} \alpha_{2p} + g_T \alpha_T \alpha_{1s},
\end{array}\right.
\eeq
where $\omega_i=\epsilon_i/\hbar$. Under the assumption of a strong THz field, we can write down the solution for $\alpha_T=\alpha_{T0} \ee^{-i\omega_T t}$. Now, with the
transformation $\alpha_{2p}\rightarrow \alpha_{2p} \ee^{-i \omega_T t}$ we obtain the system
\beq
i\hbar \frac{d}{dt}
\left( \begin{array}{c} \alpha_C \\ \alpha_{1s} \\ \alpha_{2p} \end{array} \right) =
H_{single}
\left( \begin{array}{c} \alpha_C \\ \alpha_{1s} \\ \alpha_{2p} \end{array} \right)
\eeq
where $H_{single}$ is given by (\ref{H_single}) with $\alpha_T\rightarrow\alpha_{T0}$. The result is the same as in previous sections
for obvious reasons.
Since we have shown above that in the general case there are 3 spectral lines given by $H_{single}$, it is also true, in particular,
if polaritons are in a coherent state.

\subsection{Model including pumping and losses}

\begin{figure*}
\includegraphics[width=16cm]{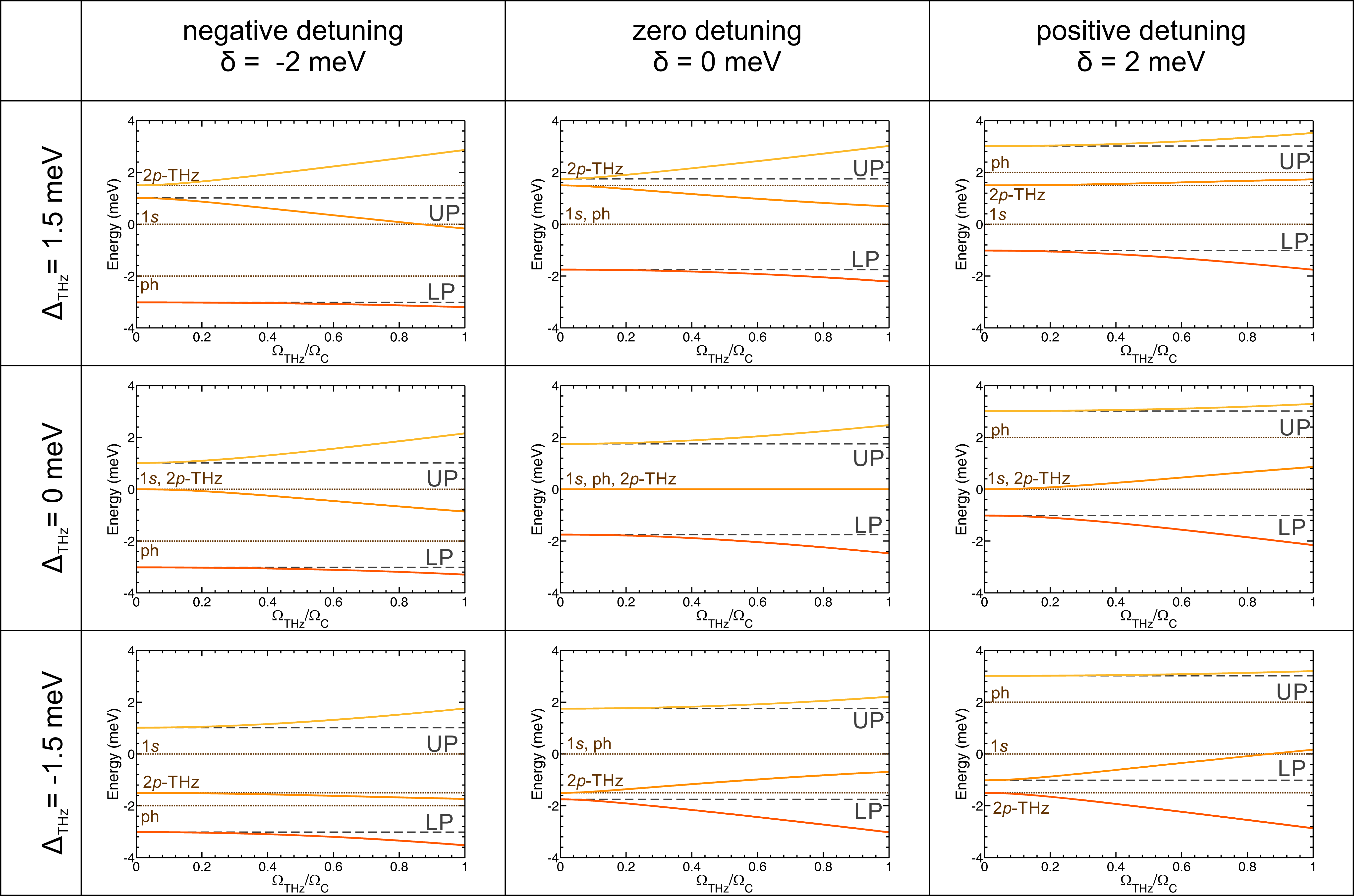}
\caption{Solution of the Hamiltonian (\ref{H_single}) for increasing THz coupling strength $\Omega_T$ from $\Omega_T$~=~0 to $\Omega_T~=~\Omega_C$~=~3.5~meV. Other parameters are as follows: $\epsilon_{1s}$~=~0~meV, $\epsilon_{2p}$~=~6.8~meV, $\Delta_{THz}=\epsilon_{2p}-\epsilon_{1s}-\epsilon_T$ is the detuning of the THz field, and $\delta=\epsilon_C-\epsilon_{1s}$ is the exciton-photon detuning.}
\label{fig:tabelka}
\end{figure*}

The above considerations allow to understand the spectrum of excitations in the conservative limit and the dependence of the coupling $\Omega_T$ on the terahertz intensity, but to obtain the full response and line broadening, both pumping and losses have to be included. Below, we describe the model that was used to obtain theoretical plots in Fig.~2 in the main text.

There are two main sources of losses in the system: the finite lifetime of cavity photons, and ionisation of excitons
due to the strong terahertz field. The ionisation rate of 1$s$ excitons is smaller than 2$p$ excitons
due to their larger binding energy. Both sources of loss can be included in the quantum description by coupling
to external optical modes (cavity losses) and to higher excited electronic states (ionisation).
External pumping by the weak probe laser can be included by coupling of the cavity mode to a classical pump field.
However, such system is no longer treatable analytically. A significant reduction of complexity can be achieved
by performing the Markovian approximation to obtain the standard form of the Master equation
with Lindblad terms for each of the sources of loss.

In the classical treatment, the problem reduces further, as pumping and losses can all be described by parameters
$\gamma_C$ for the decay rate of cavity photons due to coupling to external modes,
$\gamma_X^{1s,2p}$ for the decay rate of excitons, and $F_p=F_{p0}(t)e^{-i\omega_p t}$
for the amplitude of the probe laser field with the frequency $\omega_p$.
Considering the effect of ionisation, the value of the Keldysh parameter
$\gamma=\omega_{\rm THz}\sqrt{2 m^* E_{\rm 2p}}/e F_{\rm THz}\approx 1$, where $E_{\rm 2p}$ is the 2$p$ exciton ionisation energy and
$F_{\rm THz}$ is the peak electric field amplitude,
allows to treat the terahertz field approximately as a classical field and neglect multiphoton absorption.
In this regime~\cite{Miller_ExcitonIonization}, the decay of excitons is mainly due to
2$p$ state tunneling ionisation, and to the leading order given by an exponential dependence
$\gamma_X^{1s,2p}=\gamma_{X0}^{1s,2p} + \gamma_I^{1s,2p} e^{-\Omega_I^{1s,2p}/\Omega_T}$, where $\Omega_I$ is the tunneling ionisation threshold
and $\gamma_{X0}^{1s,2p}$ corresponds to other sources of exciton decay.

The experimental setup allows for probing the spectrum of the system with femtosecond time resolution.
Since the duration of the THz pulse is much longer
than the fs probe pulse, for each delay between the two pulses we treat the value of $\Omega_T$ as a constant
(we can treat $N_T$ as a constant since number fluctuations
in a coherent state are negligible for large occupations~\cite{CohenTannoudji_AutlerTownes}).
We also use the slowly varying envelope approximation for the fs pulse, which allows to treat $F_{p0}(t)$ as a constant
when calculating the response of the system to the probe field.
Under the assumption of a strong THz field we have $\alpha_T=\alpha_{T0} \ee^{-i\omega_T t}$ with $\alpha_{T0}$
a constant. Now, with the
transformation $\alpha_{2p}\rightarrow \alpha_{2p} \ee^{-i \omega_T t}$ we obtain the system
\beq
\left\{
\begin{array}{l}
i \dot{\alpha}_C = \omega_C \alpha_C + g_C \alpha_{1s} - i\gamma_C \alpha_C + \frac{F_{p0}}{\hbar} e^{-i\omega_p t},\\
i \dot{\alpha}_{1s} = \omega_{1s} \alpha_{1s} + g_C \alpha_{C} + g_T \alpha_{T0}^* \alpha_{2p} - i \gamma_X^{1s} \alpha_{1s},\\
i \dot{\alpha}_{2p} = (\omega_{2p} - \omega_T) \alpha_{2p} + g_T \alpha_{T0} \alpha_{1s} - i \gamma_X^{2p} \alpha_{2p}.
\end{array}\right.
\eeq
By substituting $\alpha_C=\alpha_{C0} \ee^{-i\omega_p t}$, $\alpha_{1s}=\alpha_{1s0} \ee^{-i\omega_p t}$
and $\alpha_{2p}=\alpha_{2p0} \ee^{-i\omega_p t}$ we obtain a set of linear algebraic equations that can be written in the matrix form
\begin{align}
&\left[\left(
\begin{array}{ccc}
\epsilon_C-i\frac{\hbar\gamma_C}{2} & \frac{\Omega_C^*}{2} & 0 \\
\frac{\Omega_C}{2} & \epsilon_{1s}-i\frac{\hbar\gamma_X^{1s}}{2} & \frac{\Omega_T^*}{2} \\
0 & \frac{\Omega_T}{2} & \epsilon_{2p}-\epsilon_T-i\frac{\hbar\gamma_X^{2p}}{2} \nonumber
\end{array}
\right)\right. - \\\nonumber
&-\hbar \omega_p\Bigg]
\left( \begin{array}{c} \alpha_{C0} \\ \alpha_{1s0} \\ \alpha_{2p0} \end{array} \right) =
\left( \begin{array}{c} F_{p0} \\ 0 \\ 0 \end{array} \right),
\end{align}
With pumping and losses included the spectral lines have finite width and approximately Lorentzian shape. Additionally,
polariton lines become weaker and broader with the increase of the 2$p$ ionisation as the terahertz intensity increases.
This effect is clearly visible in the experiment at large THz pulse intensities.
Additionally, we take into account the dynamical Franz-Keldysh effect~\cite{Nordstrom_DFKE}, which blue-shifts the energy of
the exciton states by an amount proportional to the terahertz field intensity,
$\epsilon_{1s,2p}(\Omega_T)=\epsilon_{1s,2p}(\Omega_T=0)+\beta_{1s,2p} \Omega_T^2$.
This effect is most clearly visible in the blueshift of the middle polariton line with increasing THz intensity.
Parameters used in our modeling are $\epsilon_C = 1.4852$~eV, $\gamma_C = 0.42$~ps$^{-1}$, $\Omega_C = 3.6$~meV, $\Omega_{T}=\Omega_{T0}e^{-(t/\tau)^2/2}$ with $\tau=7.6$~ps and $\Omega_{T0} = 3.64$~meV for THz power of 220 mW, $\delta = 0$, $\Delta = -0.70$~meV, $\Omega_I = 6.86$~meV, $\gamma_{X0}^{1s} = 0.52$ ps$^{-1}$, $\gamma_{I}^{1s} = 0.52$~ps$^{-1}$, $\gamma_{X0}^{2p} = 1.05$~ps$^{-1}$, $\gamma_{I}^{2p} = 0.78$~ps$^{-1}$, $\beta_{1s} = 0.087$~meV$^{-1}$, $\beta_{2p} = 0.175$~meV$^{-1}$.

The exciton blueshift induced by the Franz-Keldysh effect suggests that the polariton lines will move to high energies with increasing THz intensity. However, this is not the case for the middle polariton line, which becomes increasingly photonic for largest THz intensities achievable in the experiment. In result, the position of this line tends to the cavity energy $\epsilon_C$.\\

\section{Doubly dressed exciton for detuned THz energy}

\begin{figure*}
\includegraphics[width=16cm]{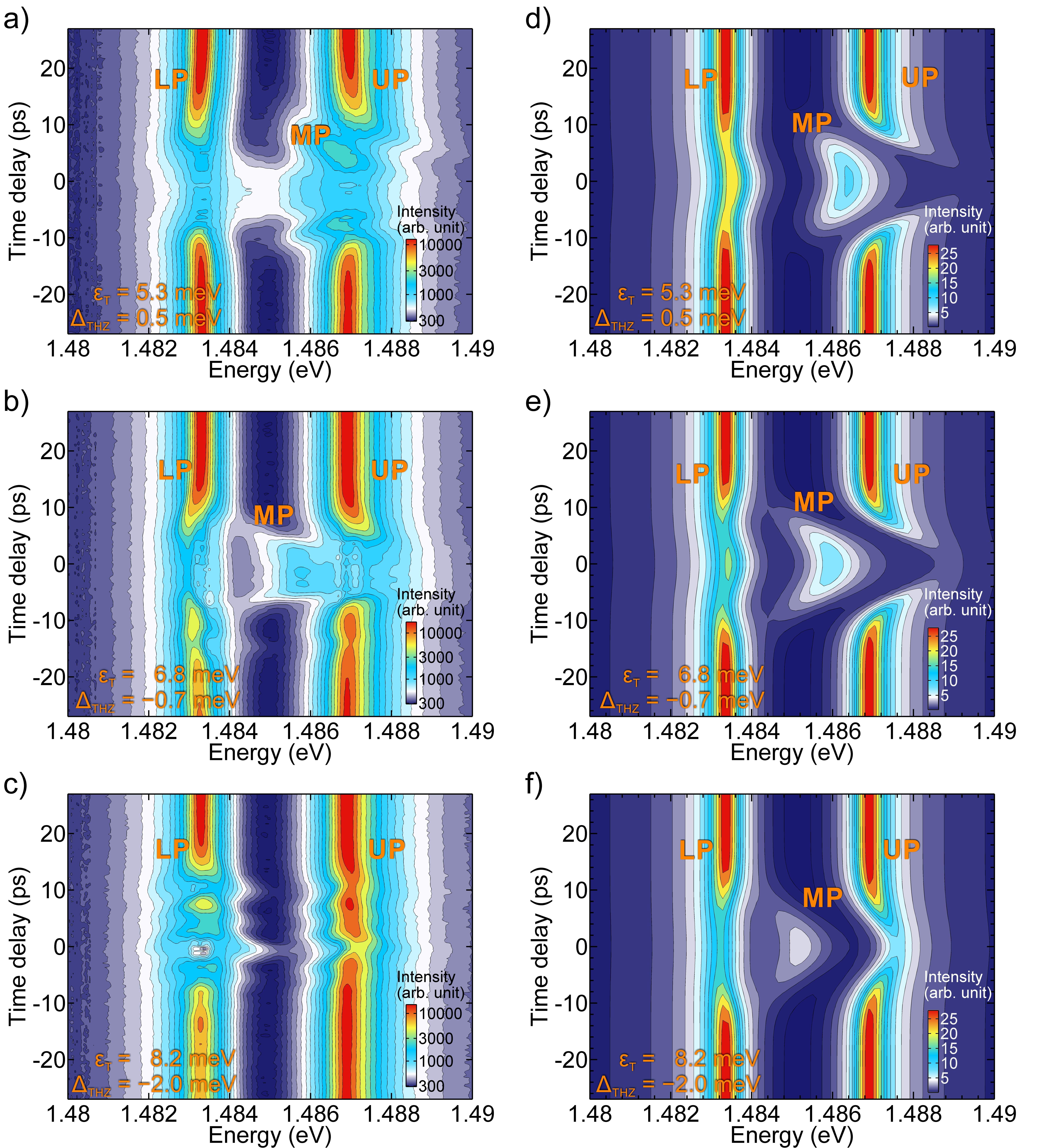}
\caption{Transmission spectra of lower (LP) and upper (UP) polaritons (at normal incidence) as a function of the delay between the NIR and THz pulses (in equidistant time intervals) for different THz photon energies: 
a) $\epsilon_T$~=~5.3~meV, i.e.~220~$\mu$m and $\Delta_{THz}$~=~0.5~meV, 
b) $\epsilon_T$~=~6.80~meV, i.e.~182~$\mu$m and $\Delta_{THz}$~=~-0.7~meV slightly negatively detuned to 1$s$--2$p$ excitonic transition and 
c) $\epsilon_T$~=~8.2~meV, i.e.~153~$\mu$m and $\Delta_{THz}$~=~-2~meV. 
d), e), f) Corresponding theoretical model. The color scale is nonlinear to enhance the low-intensity signal at zero delay time.}
\label{fig:detuning}
\end{figure*}

The possibility to change the NIR exciton - photon detuning, $\delta=\epsilon_C-\epsilon_{1s}$, by choosing the position on the sample, and the possibility to change the detuning of a THz field, $\Delta=\epsilon_{2p}-\epsilon_{1s}-\epsilon_T$, by tuning the THz laser, allows to obtain different energies of the three dressed states: LP, MP and UP. Fig.~\ref{fig:tabelka} demonstrates the doubly dressed state energies for the case of negative, zero and positive NIR and THz detunings.

Fig.~\ref{fig:detuning} illustrates the experimental results and the theoretical calculations for different photon energies of the THz beam.

For low THz energy ($\epsilon_T=5.3$~meV and $\epsilon_T=6.8$~meV) we observe almost ideal agreement between the experiment (Fig.~\ref{fig:detuning} a and b) and the theoretical model (Fig.~\ref{fig:detuning} d and e), respectively.

For the THz energy of $\epsilon_T=8.2$~meV, Fig.~\ref{fig:detuning} c and f, we do not observe the appearance of doubly dressed exciton. The signal is perturbed by the THz beam, but the middle polariton (MP) does not appear in the spectrum. We believe that the THz energy is too high to couple only the 1$s$ to 2$p$ exciton and higher excitonic states should be considered. Therefore our theoretical model does not describe this situation.

\end{document}